\documentclass[10pt,conference,letter]{IEEEtran}
\pdfoutput=1 
\usepackage{graphics}
\usepackage{bm}
\usepackage{amsmath}
\usepackage{graphicx}
\usepackage{amsmath}
\usepackage{amsfonts}
\usepackage{color}
\usepackage[]{algorithm2e}

\let\MYoriglatexcaption\caption
\renewcommand{\caption}[2][\relax]{\MYoriglatexcaption[#2]{#2}}

\newtheorem{lemma}{Lemma}

\renewcommand{\eqref}[1]{(\ref{#1})}

\begin{document}

\title{Random Access Protocols with Collision Resolution in a Noncoherent Setting}

\author{\begin{tabular}{c}
Zoran~Utkovski $^\#$$^*$, Tome~Eftimov$^*$ and Petar Popovski$^\dagger$\\
$^\#$ Faculty of Computer Science, University Goce Del\v{c}ev \v{S}tip, R. Macedonia \\
$^*$ Laboratory for Complex Systems and Networks, Macedonian
Academy of Sciences and Arts\\$^\dagger$Department of Electronic Systems, Aalborg University, Denmark
\end{tabular}
\thanks{This work was supported in part by the "Deutsche Forschungsgemeinschaft" (DFG),
via the project Li 659/13-1.}
}

\maketitle

\begin{abstract}
Wireless systems are increasingly used for Machine-Type  Communication (MTC), where the users sporadically send very short messages. In such a setting, the overhead imposed by channel estimation is substantial, thereby demanding \emph{noncoherent} communication. In this paper we consider a noncoherent setup in which users randomly access the medium to send short messages to a common receiver. We propose a transmission scheme based on Gabor frames, where each user has a dedicated codebook of $M$ possible codewords, while the codebook simultaneously serves as an ID for the user. The scheme is used as a basis for a simple protocol for collision resolution. 
\end{abstract}
\section{Introduction}

Multiple access channel (MAC) can be treated in two widely different settings~\cite{gallager}. In the \emph{ergodic} regime, the MAC capacity region is computed assuming that set of transmitting users is known in advance and does not change during many channel uses, over which each user communicates with vanishing probability of error (PoE). In the \emph{transient} regime, the users apply a random access protocol (RAP) to help the receiver learn which users are active as well as decode their data. In the transient regime there is no guarantee for vanishing PoE as the user packet have finite length. Nevertheless, the packet length in a RAP is assumed sufficiently large, such that (a) all auxiliary procedures can take place, e.g. channel estimation (b) the metadata required to run RAP is separated from the coding/decoding design of the data sent by each user through the RAP. As the message length becomes shorter, the resources required to estimate the channel become noticeable, while the separation of metadata and data becomes suboptimal. 

In this paper we consider the problem of multiple access for a set of $N$ users connected to the same receiver. Each user is activated randomly; the receiver knows $N$, but not the number $\tilde{n}$ of active users and their IDs. The model is related to the emerging scenarios of Machine-Type Communication (MTC): (1) the message of each user is very short, such that there is no dedicated metadata to detect the activity; (2) the users (devices) are simple and can neither use precoding nor invest resources to enable channel estimation at the receiver, such that the communication is noncoherent. This is relevant if the channel has a very short coherence time and the receiver jointly detects 
the set of active users as well as their data. If the number of accessing users is larger than a threshold, then the receiver detects a collision and instructs the users to retransmit using randomization, as it is common in RAPs. Our noncoherent scenario requires generalization of the notion of collision to be an event in which the receiver declares error in its attempt to jointly determine the set of active users as well as their data. We propose a transmission scheme based on \emph{Gabor frames}, where each user has a dedicated codebook of $M$ possible codewords. We introduce a decoding criterion that integrates data decoding and collision detection. Upon collision, the receiver sends feedback to initiate randomized retransmission, thus executing a collision resolution algorithm. 

In a recent related work \cite{Wunder14} the authors consider the combination of metadata to detect the activity and decode the data; yet, there are sufficient resources for channel estimation, which is not the case in our scenario with very short messages. In our scenario the metadata is extremely minimized and we do not assume that the user sends a dedicated ID, but it is identified based on the codebook that is allocated to the user. This represents rather an extreme example of integration of control and data, suited for short messages.

\section{System Model and Problem Formulation}
\label{sec:System_Model}
We consider a system with $N$ users. The system operates with slotted time. A block of $M$ slots is termed \emph{frame}.
The random access feature is captured by state
variables $\mathrm{a}_n\in\{0,1\}$ depending on whether the
$n$-th user is active $(1)$ or inactive $(0)$ in the frame. When the users are active, they may access all $M$ time slots in one frame. For simplicity, we consider the symmetric scenario where each active user transmits with the same average power $P$.

We assume block Rayleigh fading \cite{Marzetta}  \cite{lozano2008interplay}, where the channel is constant in a block
of certain length and then changes in an independent realization. 
The fading channels between the users and the common receiver remain constant within one frame, i.e. the coherence time is $T\geq M$. Each user sends a short message, which span over $K$ frames. We will treat the extreme case $K=1$, where the users communicate several bits in a $M$-slot frame. We remark that if $K>1$, one can use a concatenated code, where the inner code is applied within each frame and the outer code is a channel code that spans over a block of $K$ frames, which should ideally average the effect of fading across the frames. 
Short messages are particularly relevant in a random access setting where the users have limited power and low  processing capability, while the receiver needs to detect user activity prior to detecting user data. 
The set of active users in a frame will be termed \emph{active set} and its size follows a binomial distribution,
$\mathrm{P}_{\tilde{\mathrm{N}}}(\tilde{n})=\binom{N}{\tilde{n}}p^{\tilde{n}}(1-p)^{N-\tilde{n}}$.

We assume that the receiver does not have knowledge of the fading coefficients, the number and the identity of active users. The users and the receiver know only the statistics of the fading and the probability of activation $p$. This is referred to as \textit{non-coherent} random access channel. 
We denote by $\mathbf{x}_n
=\left[
  \mathrm{x_{n,1}}\:\mathrm{x_{n,2}}\:\cdots\:\mathrm{x_{n,M}}\right]^{\mathrm{T}}$ the transmit vector of user $n$ in one frame.  The received signal vector
$\mathbf{y}=\left[
  \mathrm{y_1}\:\mathrm{y_2}\:\cdots\:\mathrm{y_M}\right]^{\mathrm{T}}$ is
\begin{equation}
\mathbf{y}=\sum_{n=1}^N\mathbf{x}_n\mathrm{a_n}\mathrm{h_n}+\mathbf{w},
\label{eq:system_model}
\end{equation}
where $\mathrm{h_n}$ is the channel coefficient of user $n$ and $\mathbf{w}=\left[\mathrm{w_1}\:
\mathrm{w_2}\:\cdots\:\mathrm{w_M}\right]^{\mathrm{T}}$ is the noise vector. The channel coefficients $\mathrm{h_n}$ and the elements of $\mathbf{w}$ are circular complex
Gaussian with zero mean and variance 1. 
It is convenient to write (\ref{eq:system_model}) as
\begin{equation}
\mathbf{y}=\mathbf{X}\mathbf{A}\mathbf{h}+\mathbf{w},
\label{eq:system_model_matrix}
\end{equation}
where the columns of $\mathbf{X}=\left[
  \mathbf{x}_1\:\mathbf{x}_2\:\cdots\:
  \mathbf{x}_N\right]^{\mathrm{T}}\in \mathbb{C}^{M\times N}$ 
represent the users, and $\mathbf{A}$ is a
diagonal matrix to represent the random
activity in the frame, $\mathbf{a}=\left[
  \mathrm{a_1}\:\mathrm{a_2}\:\cdots\:\mathrm{a_N}\right]$. We index the $2^N$ possible
realizations of $\mathbf{a}$ as $a_{\tilde{n},i}$ where $\tilde{n}$ refers to the
number of active users in the frame and $i$ to the particular active set. 
For fixed $\tilde{\mathrm{N}}=\tilde{n}$, there are $\binom{N}{\tilde{n}}$ distinct
active sets of $\tilde{n}$ users, each of them occurring
with probability $\mathrm{P}_{\mathbf{a}}(a_{\tilde{n},i})=p^{\tilde{n}}(1-p)^{N-\tilde{n}}$. Alternatively, we can write
\begin{equation}
\mathbf{y}=\tilde{\mathbf{X}}\tilde{\mathbf{h}}+\mathbf{w},
\label{eq:alternative_model}
\end{equation}
where $\tilde{\mathbf{X}}$ is
obtained from $\mathbf{X}\mathbf{A}$ by simply deleting
the all-zero columns which correspond to the inactive users, and $\tilde{\mathbf{h}}$ is a vector which
contains the corresponding channel coefficients. 

\section{Non-coherent Random Access Communication}
\label{sec:Communication}
\subsection{Preliminaries}
The received vector $\mathbf{y}$ in one frame
represents a linear combination of the signal vectors from the active users. Therefore $\mathbf{y}$ lies in the
linear subspace of $\mathbb{C}^M$ spanned by the transmit
vectors of the active users. This set up conforms to the geometric interpretation of the communication over non-coherent MIMO channels \cite{Tse}. The users convey information by subspaces, such that even without the knowledge
of the active set and the channel coefficients, the receiver can resolve the active users along with their messages, provided that the subspaces spanned by the transmit vectors of the active users are distinct. This
observation is the essence of our framework.

The random access channel with users' states known at the receiver can be mapped to a multi-receiver channel, by introducing one auxiliary
receiver per each channel state, see \cite{Minero}. However, as in our non-coherent setting the receiver does not know the active set size in advance, the model in \cite{Minero} is no longer valid in the presented form. In our case, the decoder has to search over the codeword combinations of all possible active sets of size $0$ to $N$, $\tilde{n}=0,\ldots,N$. 
Let 
$\mathcal{C}_n=\left\{\mathbf{c}_n^{(1)},
\mathbf{c}_n^{(2)},\ldots,\mathbf{c}_n^{(S_n)}\right\}$ be the codebook of user $n$,
where the codewords $\mathbf{c}_n^{(j)}\subset \mathbb{C}^M$, and $S_n\doteq\vert\mathcal{C}_n\vert$ is the cardinality of $\mathcal{C}_n$.
Let us define by $\mathcal{X}_{\tilde{n},i}$ the set which contains all $M\times \tilde{n}$ matrices  obtained by concatenation of the codewords of the $i$-th combination of $\tilde{\mathrm{N}}=\tilde{n}$ active users (out of $N$). We note that, according to this notation, $\mathcal{X}_{1,i}$ corresponds to the codebook of user $i$ itself, $\mathcal{X}_i\equiv\mathcal{C}_i$, $i=1,\ldots, N$. The trivial set $\mathcal{X}_{0,1}=\{0\}$ corresponds to the case when no user is active in the frame. 
As result, the cardinality of $\mathcal{X}_{\tilde{n},i}$
is $\vert \mathcal{X}_{\tilde{n},i} \vert=\prod_{n=1}^{N}S_n\bm{a}_{\tilde{n},i}^{(n)}$, 
where $\bm{a}_{\tilde{n},i}^{(n)}=\left(a_{\tilde{n},i}^{(1)},\ldots,a_{\tilde{n},i}^{(N)}\right)$ is a vector with  $a_{\tilde{n},i}^{(n)}\in\{0,1\}$ that describes the user activity: the non-zero elements correspond to the $i$-th combination of $\tilde{n}$ out of $N$ active users. In addition, we define $\mathcal{X}_{\tilde{n}}$ as the set of all ``effective' codewords when $\tilde{n}$ users are active. By definition,  
$\mathcal{X}_{\tilde{n}}=\bigcup_{i=1}^{\binom{N}{\tilde{n}}} \mathcal{X}_{\tilde{n},i}$. 
Finally, the "effective" codebook of the random access system is defined as 
$\mathcal{X}=\bigcup_{\tilde{n}=0}^N \mathcal{X}_{\tilde{n}}$, and is
used by the receiver in the decoding process.
\begin{figure}[!t]
\centering
\includegraphics[width=79mm]{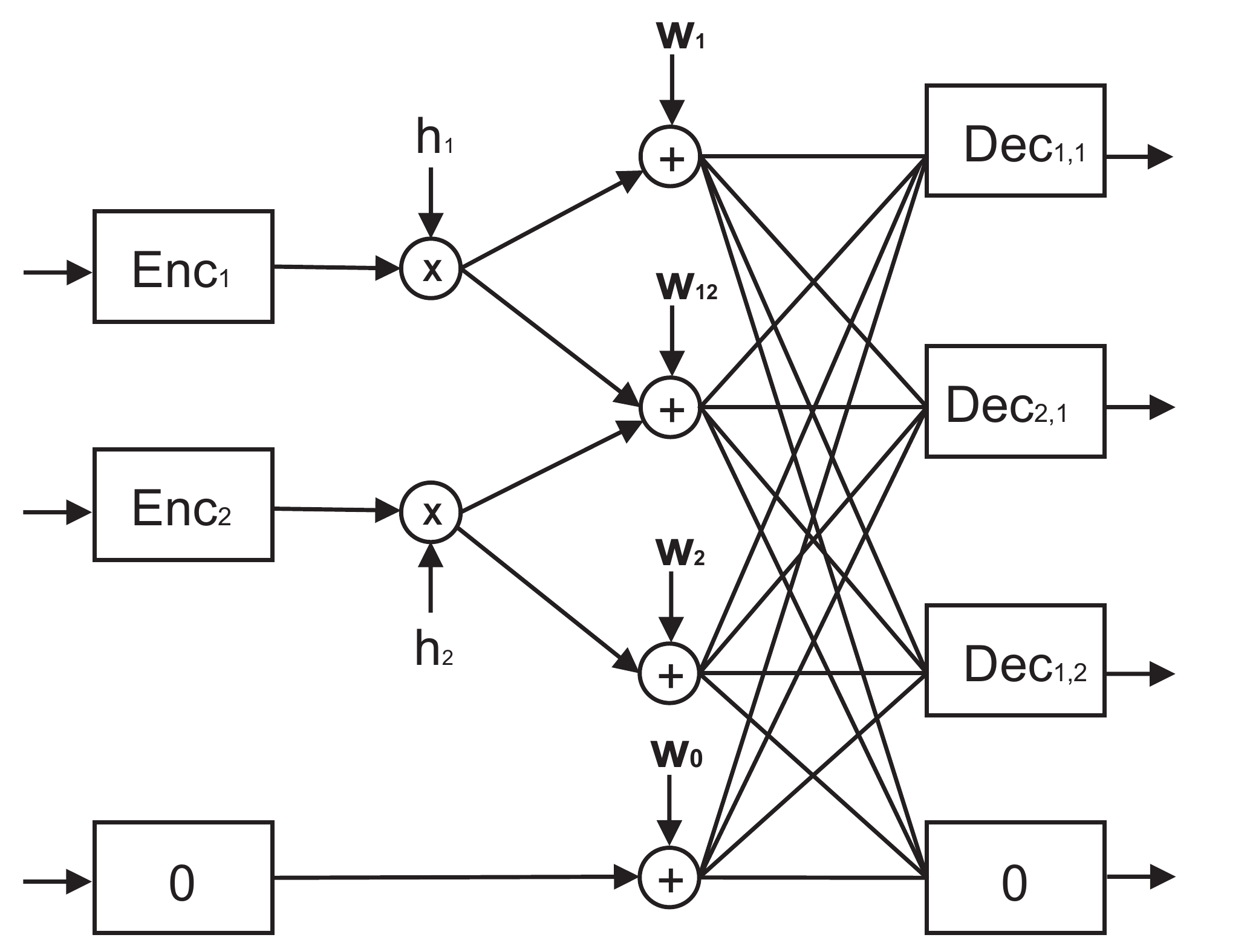}
\caption{Equivalent representation of the two-user non-coherent random access channel, with active set and channel realizations unknown to the receiver.} 
\label{fig:system_model_unknown_states}
\end{figure}
As illustration (see Fig.\ref{fig:system_model_unknown_states}), we address the case of two users with codebooks $\mathcal{C}_1$ and $\mathcal{C}_2$ respectively. Due to the nature of the random access , the receiver ``sees'' a codebook $\mathcal{X}=\left\{\mathcal{X}_{1,1}, \mathcal{X}_{1,2}, \mathcal{X}_{2,1}, \bm{0}\right\}$, where $\mathcal{X}_{1,1}\equiv \mathcal{C}_1$, $\mathcal{X}_{1,2}\equiv \mathcal{C}_2$, $\mathcal{X}_{2,1}\in \mathbb{C}^{M\times 2}$ corresponds to both users being active and $\bm{0}$ is the ``zero'' codeword when the active set is empty. It should be noted that this model is reminiscent to the model used in \cite{LuoEphremides}, where ``no transmission'' by the user is a valid data symbol, which unifies the framework of detection and data decoding. 

\subsection{Decoding} 
Having the received vector $\mathbf{y}$, the
receiver should decide on the most probable codeword of the effective codebook $\mathcal{X}$. 
From (\ref{eq:alternative_model}), the MAP decoding rule can be derived as
\begin{small}
\begin{align}
\hat{\tilde{\bm{X}}}=\arg\max_{\tilde{\bm{X}}\in\mathcal{X}}\frac{\exp\left\{\mathrm{tr}\left[\mathbf{y}^\mathrm{H}\tilde{\bm{V}}\left(\bm{I}_M+\frac{\tilde{\bm{\Lambda}}^{-1}}{\rho M}\right)^{-1}\tilde{\bm{V}}^\mathrm{H}\mathbf{y}\right]\right\}}{\pi^M\det\left(\bm{I}_{\tilde{n}}+\rho M \tilde{\bm{\Lambda}}\right)}
\cdot P_{\tilde{\mathbf{X}}}(\tilde{\bm{X}}),
\label{eq:MAP_decoding}
\end{align}
\end{small}\noindent 
where $\tilde{\bm{X}}=\tilde{\bm{V}}\tilde{\bm{\Lambda}}^{1/2}\tilde{\bm{U}}$ is the svd decomposition of $\tilde{\bm{X}}$, with $\tilde{\bm{V}}$ being $M\times \tilde{n}$ unitary, $\tilde{\bm{\Lambda}}$ being $\tilde{n}\times\tilde{n}$ diagonal with non-negative entries (the eigenvalues of $\tilde{\bm{X}}^\mathrm{H}\tilde{\bm{X}}$), and $\tilde{\bm{U}}$ being $\tilde{n}\times \tilde{n}$ unitary. $\tilde{\bm{V}}$ can also be interpreted as a basis for the subspace spanned by $\tilde{\bm{X}}$. 
According to (\ref{eq:MAP_decoding}), the receiver performs a search over all $\mathcal{X}_{\tilde{n}}$, $\tilde{n}=0,\ldots,N$. Having $\tilde{\bm{X}}$, the receiver determines simultaneously the active set 
and the codewords from the users' codebooks which produced the effective codeword $\tilde{\bm{X}}$.  

In non-coherent point-to-point MIMO channels, the ML decoding rule for the capacity-achieving input signals in high SNR suggests to convey information by using linear subspaces with \textit{chordal distance} as a relevant construction metric \cite{Marzetta}, \cite{Tse}. However, the MAP decoding rule (\ref{eq:MAP_decoding}) does not suggest a straightforward coding strategy as it does not offer a simple geometric interpretation of the decision-relevant metric. 
Nevertheless, information through the random access channel can still be conveyed by subspaces, as implicitly suggested by the decoding rule. Indeed, if we assume that the receiver knows the number of active users $\tilde{n}$, the decoding rule becomes
\begin{align}
\hat{\tilde{\bm{X}}}=\arg\max_{\tilde{\bm{X}}\in\mathcal{X}_{\tilde{n}}}\Vert\mathbf{y}^\mathrm{H}\tilde{\bm{V}}\left(\bm{I}_M+\frac{1}{\rho M}\tilde{\bm{\Lambda}}^{-1}\right)^{-1/2}\Vert_{\mathrm{F}}^2,
\label{eq:ML_decoding_simple}
\end{align} 
which can be interpreted as projection of $\bm{y}$ on the linear subspace $\tilde{\bm{V}}$ spanned by $\tilde{\bm{X}}$, where the projection is corrected by the term $\left(\bm{I}_M+\frac{1}{\rho M}\tilde{\bm{\Lambda}}^{-1}\right)^{-1/2}$. In point-to-point MIMO with unitary space-time modulation this term is an identity matrix \cite{Marzetta}, reducing (\ref{eq:ML_decoding_simple}) to the known ML-decoding rule.

When the receiver knows $\tilde{n}$, then (\ref{eq:ML_decoding_simple}) implies that the problem of detecting the actual set of active 
users and their codewords
translates to a problem of linear subspace detection in the vector space
$\mathbb{C}^M$. Decoding is possible as long as the subspace spanned by the effective codeword $\tilde{\bm{X}}$, uniquely defined by $\tilde{\bm{V}}$, can be revealed without ambiguity from $\mathbf{y}$. Here \emph{collision} is the event when the decoding fails for the reason that 
combination of codewords from the active set does not produce a unique subspace. As the subspaces corresponding to the elements of the effective codebook $\mathcal{X}$ are of different dimensions, spanning from $0$ to $M$, the relevant coding space is related to the union of the Grassmannians, $\bigcup_{m=0}^M
\mathcal{G}_{M,m}^{\mathbb{C}}$, where the set $\mathcal{G}_{M,m}^{\mathbb{C}}$ denotes the collection of $m$-dimensional linear subspaces of $\mathbb{C}^M$, also known as Grassmann manifolds. 

\section{Coding for Noncoherent Random Access}
 
\subsection{Codes from Gabor frames}


In the following we present a construction that is based on Gabor frames \cite{Strohmer, Calderbank}. 
For a prime $M\geq 5$, there is a Gabor frame $\left\{\bm{g}_{k,l}\right\}_{k,l=0}^{M-1}$ with vectors in $\mathbb{C}^M$ given by
\begin{equation}
\bm{g}_{k,l}=\bm{g}((m-k)\:\mathrm{mod}\:M)e^{2\pi ilm/M},\:k,l=0,\ldots, M-1,
\label{eq:Gabor_frame}
\end{equation} 
where
 $\bm{g}(m)=e^{2\pi im^3/M}$, $m=0,1,\ldots,M-1$. 
The set (\ref{eq:Gabor_frame}) represents a Grassmannian frame, a selection of lines going through the origin of $\mathbb{C}^M$ (also pointed out by the authors in \cite{Strohmer}). It has $M^2$ elements with corresponding scalar products
\begin{equation}
\vert \bm{g}_{k,l}\bm{g}^{\mathrm{H}}_{k',l'}\vert \in \left\{0,1/\sqrt{M}\right\},\:\forall (k,l)\neq (k',l').
\label{eq:frame_correlation}
\end{equation}
This frame is actually a union of $M$ orthonormal bases of $\mathbb{C}^M$ and  
by assigning one basis to each user we can accommodate $N=M$ users 
\footnote{In fact, one can add the standard orthonormal basis to the frame without changing the maximal frame correlation $1/\sqrt{M}$.},  
with a data rate of up to $(\log_2 M)/M$ bits/channel use per active user if there is no collision. 

Formally, the codebooks assigned to the users are defined as $\mathcal{C}_1=\left\{\bm{g}_{0,l}\right\}_{l=0}^{M-1}$, $\mathcal{C}_2=\left\{\bm{g}_{1,l}\right\}_{l=0}^{M-1}$, $\ldots,$ $\mathcal{C}_M=\left\{\bm{g}_{M-1,l}\right\}_{l=0}^{M-1}$. According to Section \ref{sec:Communication}, the uncertainty in the number of active users $\tilde{\mathrm{N}}$ gives rise to the effective codebook
$\mathcal{X}=\bigcup _{\tilde{n}=0}^M \mathcal{X}_{\tilde{n}}$, 
where $\mathcal{X}_{\tilde{n}}$ itself is the union 
$\mathcal{X}_{\tilde{n}}=\bigcup_{i=1}^{\binom{N}{\tilde{n}}} \mathcal{X}_{\tilde{n},i}$.
Hence, in the presented construction, the size of the effective codebook, as seen by the receiver, is  
$\vert \mathcal{X} \vert=\sum_{\tilde{n}=0}^M \binom{M}{\tilde{n}}M^{\tilde{n}}$.
\subsection{Decoding and collision resolution}
From (\ref{eq:frame_correlation}) one can show that (except for the case $\tilde{n}=0$), the ordered eigenvalues of 
$\tilde{\bm{X}}^\mathrm{H}\tilde{\bm{X}}$ ($\forall \tilde{\bm{X}}\in\mathcal{X}$) are 
\begin{align}
\tilde{\lambda}_1=1+(\tilde{n}-1)\sqrt{M};\:
\tilde{\lambda}_j=1-\sqrt{M},\: j=2,\ldots,\tilde{n}.
\label{eq:Gabor_eigenvalues}
\end{align}
The eigenvalues are plugged in the MAP decoding rule (\ref{eq:MAP_decoding}).   
While in the general case of unknown $\tilde{\mathrm{N}}$ the MAP decoder has to account for $\tilde{\Lambda}$, for given $\tilde{\mathrm{N}}=\tilde{n}$, the rule (\ref{eq:MAP_decoding}) becomes
\begin{small}
\begin{align}
\hat{\tilde{\bm{V}}}=\arg\max_{\tilde{\bm{X}}\in\mathcal{X}_{\tilde{n}}}\Vert\mathbf{y}^\mathrm{H}\tilde{\bm{V}}\Vert_{F}^2 P_{\tilde{\mathbf{X}}\vert \tilde{\mathrm{N}}}(\tilde{\bm{X}}|\tilde{n})=\arg\max_{\tilde{\bm{X}}\in\mathcal{X}_{\tilde{n}}}\Vert\mathbf{y}^\mathrm{H}\tilde{\bm{V}}\Vert_{F}^2,
\label{eq:Decoding_Gabor}
\end{align}
\end{small}\noindent 
since $P_{\tilde{\mathbf{X}}\vert \tilde{\mathrm{N}}}=1/\vert \mathcal{X}_n\vert$, $\forall \tilde{\bm{X}}\in \mathcal{X}_{\tilde{n}}$. 
As $\tilde{\bm{V}}$ is a matrix representative of the subspace spanned by the effective codeword $\tilde{\bm{X}}$, the decoding rule suggests that the decoder can resolve the different codewords $\tilde{\bm{X}}$ as long as the subspaces $\tilde{\bm{V}}$ corresponding to different $\tilde{\bm{X}}$ are \textit{distinct} in terms of their \textit{chordal distance}. For two subspaces $\bm{\Phi}, \bm{\Psi}\in \mathcal{G}_{M,m}^\mathbb{C}$, their chordal distance is related to the principal angles between the subspaces. Let $\bm{\Psi}^\mathrm{H}\bm{\Phi}=\mathbf{U}\mathbf{\Sigma} \mathbf{V}^\mathrm{H}, \mathbf{U},\mathbf{\Sigma},\mathbf{V}\in\mathbb{C}^{m\times m}$ be the singular value decomposition of $\mathbf{\mathbf{\Psi}}^\mathrm{H}\mathbf{\mathbf{\Phi}}$, with $\mathbf{\Sigma}$ being a diagonal matrix of the singular values $\sigma_1,\ldots,\sigma_m$. The $m$ principle angles between $\bm{\mathbf{\Phi}}$ and $\bm{\mathbf{\Psi}}$ are
$\theta_i=\mathrm{acos}\:\sigma_i$, giving rise to the chordal distance $
\mathrm{d}(\bm{\Phi},\bm{\Psi})=\sqrt{\sum_{i=1}^m\sin^2 \theta_i}$.

\begin{lemma}
\label{lemma_1}
$\forall \bm{X},\bm{Y}\in \mathcal{X}_{\tilde{n}}$ where $\bm{X}\neq\bm{Y}$ and $\tilde{n}=1,\ldots,\lfloor \frac{M}{2} \rfloor$, $d(\bm{\Phi},\bm{\Psi})>0$, i~e. distinct codewords from $\mathcal{X}_{\tilde{n}}$ correspond to distinct subspaces.\end{lemma} 
\begin{IEEEproof}The proof is omitted due to the limited room. 
\end{IEEEproof}
Lemma \ref{lemma_1} implies that, in absence of noise, the receiver can decode any combination of codewords without ambiguity if $\tilde{n}\leq \lfloor M/2 \rfloor$, i.~e. $\lfloor M/2 \rfloor$ users can be resolved in the system. This observation motivates a simple suboptimal decoding method where at first the decoder uses 
$\bm{y}$ to estimate if the number of active users is greater than $\lfloor M/2 \rfloor$. This check can be performed based on the statistics of $\mathrm{z}=\mathrm{tr}\left[\mathbf{y}\mathbf{y}^\mathrm{H}\right]$
\begin{align}
\mathrm{z}&=
\mathrm{tr}\left[\tilde{\mathbf{X}}^\mathrm{H}\tilde{\mathbf{X}}\mathbf{h}\mathbf{h}^\mathrm{H}\right]+2\Re\left\{\mathrm{tr}\left[\mathbf{X}\mathbf{h}\mathbf{w}^\mathrm{H}\right]\right\}+\mathrm{tr}\left[\mathbf{w}\mathbf{w}^\mathrm{H}\right].
\label{eq:Trace_received}
\end{align} 
The estimate is obtained by setting a threshold $t_z$ and performing the comparison $\mathrm{z}>t_z$. The selection of the $t_z$ is related to the probability of the error events, $P\left(\mathrm{z}>t_z|\tilde{\mathrm{N}}\leq \lfloor M/2 \rfloor \right)$ and $P\left(\mathrm{z}\leq t_z|\tilde{\mathrm{N}}>\lfloor M/2 \rfloor \right)$. The threshold should be set such that $\frac{P\left(z\leq t_z|\tilde{\mathrm{N}}>\lfloor M/2 \rfloor \right)}{P\left(z>t_z|\tilde{\mathrm{N}}\leq \lfloor M/2 \rfloor \right)}\leq1$ since the error event of having a false collision is less severe than the event of having a collision which is not recognised\footnote{A false collision may be resolved in the collision resolution phase, but an unrecognised collision will certainly lead to a frame error.}. 
If $z\leq t_z$, i.~e. $\tilde{\mathrm{N}}
\leq \lfloor M/2 \rfloor$, in the second step the receiver decodes based on (\ref{eq:MAP_decoding}). The decoding can be further simplified if the receiver first estimates $\tilde{\mathrm{N}}$ (given that $\tilde{\mathrm{N}}\leq \lfloor M/2 \rfloor$) and then decodes based on (\ref{eq:Decoding_Gabor}). 

If $z>t_z$ ($\tilde{\mathrm{N}}\leq \lfloor M/2 \rfloor$), the receiver observes a collision, after which it obtains an  estimate $\hat{\tilde{\mathrm{N}}}$ and announces collision, informing the users about $\hat{\tilde{\mathrm{N}}}$. In the next frame of length $M$ no new users are allowed, while the active users will remain active with probability $p_c$, where $p_c$
is adjusted to $\hat{\tilde{\mathrm{N}}}$. The choice $p_c\leq\frac{\lfloor M/2 \rfloor}{\tilde{\mathrm{N}}}$ ensures an average number of active users not greater than $\lfloor M/2 \rfloor$ and decreases the chance of repeated collision.
The proposed algorithm is summarized as \\\\

\vspace{-24pt}

\begin{algorithm}
{\small
1) Users transmit data with probability $p$ in one frame.\\
2) Receiver calculates $z=\mathrm{tr}\left[\mathbf{y}\mathbf{y}^\mathrm{H}\right]$ and threshold $t_z$.\\
3) \eIf{$ z\leq t_Z$}{
	- Estimate $\tilde{\mathrm{N}}$ and decode $\bm{y}$ based on (\ref{eq:Decoding_Gabor});\\
- If no estimate of $\tilde{\mathrm{N}}$, decode $\bm{y}$ based on (\ref{eq:MAP_decoding});\
}{
 - Estimate $\tilde{\mathrm{N}}$ and announce collision;\\
- No new users are allowed in the system;\\
- Active users retransmit with probability $p_c$;\\
- Repeat 2);\\ 
\eIf{$ z\leq t_Z$}{
- Go to 3);
}{
- Announce error and go to 1).
}
}}
\caption{Decoding with collision resolution}
\end{algorithm}     

\section{Results and Discussion}
\begin{figure}[t]
\centering
\includegraphics[width=89mm]{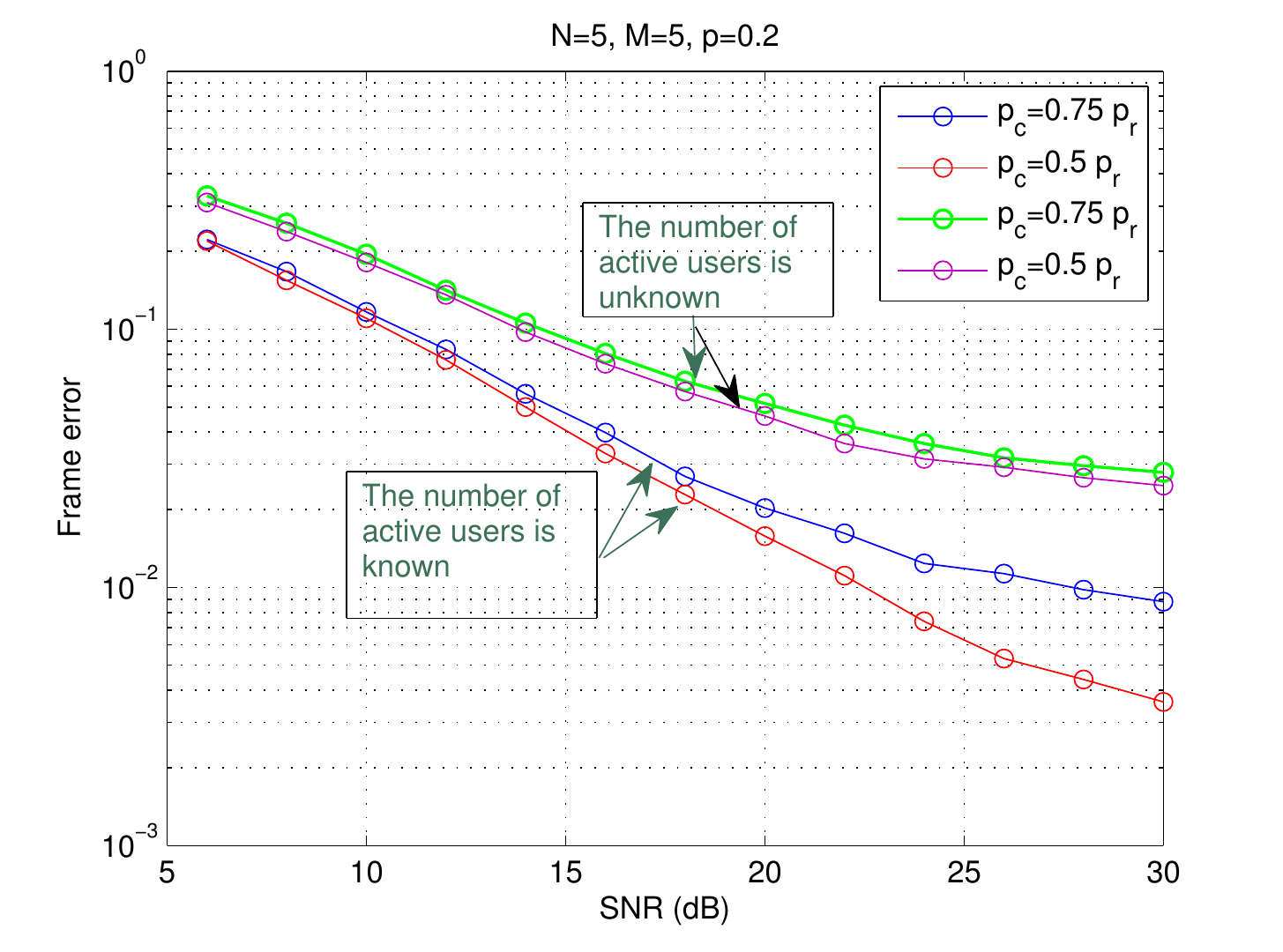}
\caption{Frame error for different probabilities of retransmission $p_c$.}
\label{fig:p_a=0.2}
\end{figure}

\begin{figure}[!t]
\centering
\includegraphics[width=89mm]{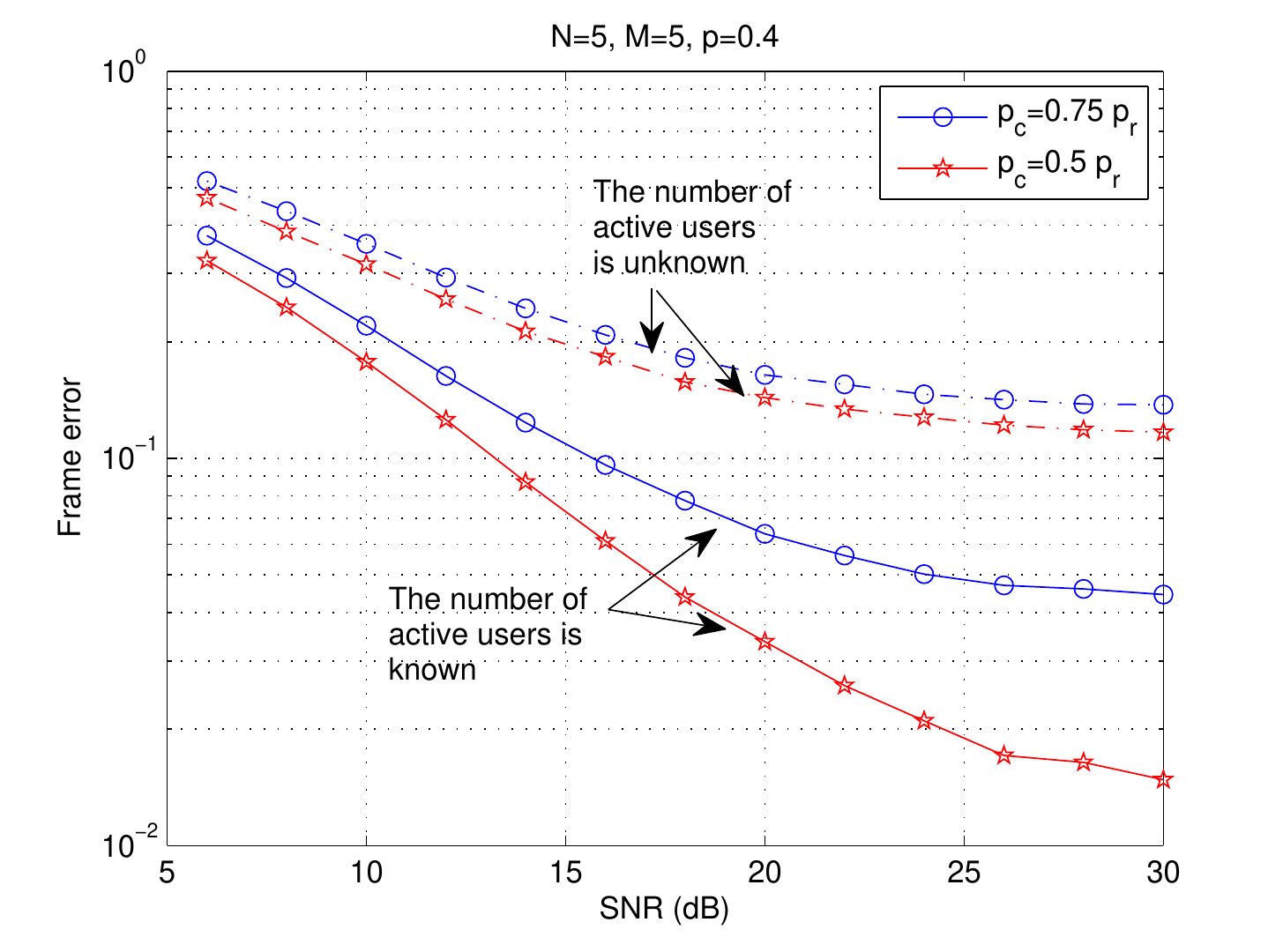}
\caption{Frame error for different probabilities of retransmission $p_c$.}
\label{fig:p_a=0.4}
\end{figure}

We evaluate the performance of the proposed scheme, where joint detection and collision resolution follows the simplified decoding procedure. As performance benchmark (upper bound) we take the case when the number of active users is known to the receiver. The results for a system with $N=5$ users and activation probabilities $p=0.2$ and $p=0.4$ are illustrated in Fig.~\ref{fig:p_a=0.2} and Fig.~\ref{fig:p_a=0.4}. 
The result shows that the error performance is dominated by the probabilities that a collision goes unrecognised and that a collision is not resolved in the collision resolution phase. While the second effect is related to $p$, the first one is largely due to the unknown amplitudes of the channel realizations. 
Our scheme is tailored to allow for communication in the most extreme example of block Rayleigh fading and short messages (of length $M$ time slots), when both the fading and the active set are unknown. The performance is expected to improve significantly in practical scenarios where the fading amplitude changes relatively slow compared to the channel phase. To improve the performance one can also resort to the (rather complicated) MAP decoding (\ref{eq:MAP_decoding}), or to a more sophisticated estimation of the number of active users based on the statistics of $\mathbf{y}$. In addition, employing multiple antennas at the receiver would improve the estimation of the number of users, and decrease the pairwise PoE. 

Although the presented version of our scheme is restricted in number of users and the throughput, it provides \textit{reliable communication} of short messages, when the block fading and the active set are unknown.  The scheme integrates the PHY/MAC layers, thereby saving the overhead that would significantly affect the short messages. Although the number of users can be increased by increasing the frame length $M$, there is a restriction imposed by the coherence time $T$ of the channel ($M\leq T$). This is a fundamental limit to \textit{any} scheme, also a coherent one, that operates over block fading since the acquisition of channel knowledge requires overhead in time slots that is in the order of the number of active users. Even if $T$ is long, the sporadic access of the users makes it ineffective to have extensive channel estimation followed by channel coding over long blocks.

If the users' messages span over $K>1$ frames, one can dedicate one frame for training, such that the the Gabor frames are used as "signatures"/pilots, instead of data signals. In this way the scheme can accommodate $N=M(M+1)$ users but, as with orthogonal pilots, not more than $\tilde{n}=M$ active users can be simultaneously resolved from the training matrix. The training period is followed by coherent transmission, once the receiver estimated the channels and the active set, based on the users' signatures. This is a subject of our future work.

\section{Acknowledgement} 
The research presented in this paper was supported by the Danish
Council for Independent Research (Det Frie Forskningsr\aa d), grant no.
11-105159 "Dependable Wireless Bits for Machine-to-Machine (M2M)
Communications", and by the German Research Society (Deutsche Forschungsgemeinschaft-DFG),
via the project Li 659/13-1 "Non-coherent Communication in Future Wireless Networks".
\bibliographystyle{IEEEtran}
\bibliography{Bibliography}

\bibliographystyle{IEEEtran}

\end{document}